\documentclass{pazh1}
\usepackage{graphicx}
\usepackage{latexsym}

\def\Ha{\hbox{H$_\alpha$\,}}
\def\Hb{\hbox{H$_\beta$\,}}

\begin{document}

\title{\bf Star Formation History in the Central Region of the Barred Galaxy NGC 7177}

\author{\bf O. K. Sil$'$chenko$^1$ and A. A. Smirnova$^2$}

\institute{
{\it $^1$ Sternberg Astronomical Institute, Universitetskii pr. 13, Moscow, 119992 Russia}\\
{\it $^2$ Special Astrophysical Observatory, Russian Academy of Sciences,
Nizhnii Arkhyz, Karachai-Cherkessian Republic, 369167 Russia}}
%\vspace{2mm}
\offprints{O. K. Sil$'$chenko, \email{olga@sai.msu.su}}
\date{}

\titlerunning{Star Formation History in NGC 7177}

\authorrunning{Sil$'$chenko and Smirnova}

\abstract{—Using the method of integral-field (3D) spectroscopy, we have investigated the kinematics and
distribution of the gas and stars at the center of the early-type spiral galaxy  with a medium scale
bar NGC 7177 as well as the change in the mean age of the stellar population along the radius. A classical
picture of radial gas inflow to the galactic center along the shock fronts delineated by dust concentration at
the leading edges of the bar has been revealed. The gas inflow is observed down to a radius R = 1$''$.5 –- 2$''$, where the gas flows at the inner Lindblad resonance concentrate in an azimuthally highly inhomogeneous
nuclear star formation ring. The bar in NGC 7177 is shown to be thick in z coordinate; basically, it has
already turned into a pseudo-bulge as a result of secular dynamical evolution. The mean stellar age inside
the star formation ring, in the galactic nucleus, is old, $\sim$10 Gyr. Outside, at a distance R = 6$''$-–8$''$ from the
nucleus, the mean age of the stellar population is $\sim$2 Gyr. If we agree that the bar in NGC 7177 is old, then,
obviously, the star formation ring has migrated radially inward in the last 1–-2 Gyr, in accordance with the
predictions of some dynamical models.
\vspace{0.5cm}

}

%\noindent
%{\bf Key words:\/} galactic nuclei, galactic structure, galactic evolution.
%{\bf PACS:\/} 98.52.Lp, 98.58.Ay, 98.62.Ai, 98.62.Hr, 98.62.Js, 98.62.Lv, 98.62.Bj
%\vfill
%\noindent\rule{8cm}{1pt}\\
%DOI: 10.1134/S1063773710050026
\clearpage
\maketitle

\section*{INTRODUCTION}

Now the understanding has matured among the researchers of galactic evolution that slow
dynamical processes can play a very important role in galaxy shaping, because
they can change strongly the global morphology of galaxies on timescales of 1 –- 5 Gyr.
Such slow processes have been called `secular evolution'. They include both the internal instabilities of dynamically cold stellar disks and the processes related to the
gravitational interaction with neighboring galaxies and to the hydrodynamical
interaction with the intergalactic medium. A good comprehensive review of the present
view on secular galactic evolution has been published by Kormendy and Kennicutt (2004).
 Interestingly, most of the secular evolution mechanisms redistribute the matter in galaxies
in such a way that the gas from the outer galactic disk regions flows into the central
region and star formation can and must take place there, leading to the formation of
new stellar subsystems: pseudobulges, chemically decoupled stellar nuclei, young
ring structures, etc.

NGC 7177 is a medium-luminosity Sb-type spiral galaxy with a well-developed medium-sized
bar (Fig. 1). The nucleus of NGC 7177 is a weak LINER (Misselt et al. 1999). The
parameters of this galaxy make it a convenient object for investigation, which is also
confirmed by the fact that NGC 7177 has been repeatedly included in the samples of both photometric and kinematical surveys. Photometric information about the structure of NGC 7177
can be found in de Jong (1996). Analyzing the galaxy's CCD images in several color bands and basing on the `standard' exponential disk plus de Vaucouleurs' bulge model, de Jong (1996)
discovered there a compact bulge (r$_e$ = 7$''$), a small dense stellar disk
(h = 13$''$ (K \textbf{band})-- 11$''$ (B \textbf{band}) or about 1.5 kpc) with
a surface brightness that is approximately triple the standard one
($\mu_{0,B}$ = 20$^m$.46 arcsec$^{-2}$), and a bar with a radius of 13$''$ oriented
at a position angle of 13$^{\circ}$ (K band) –- 18$^{\circ}$ (B band). Erwin (2005)
refined the photometric parameters of NGC 7177: the bar radius is 10$''$ -– 11$''$
and the disk orientation angles estimated from the shape of its outer isophotes by
assuming the disk to be circular are: the inclination 48$^{\circ}$ and the position angle
of the line of nodes PA$_0$ = 83$^{\circ}$. Erwin et al. (2008) analyzed the radial surface brightness profile for NGC 7177 basing on a deeper R-band image without obsessing about the concept of single-scale exponential disks. They discovered another outer stellar disk that
begins to dominate in the galactic structure at a distance of 89$''$ (about 7 kpc) from the center; this disk has a large scalelength, of 84$''$ (about 7 kpc), and a low surface
brightness, $\mu_{0,R}$ = 23$^m$.5 arcsec$^{-2}$.

\begin{figure}
\includegraphics[width=\hsize]{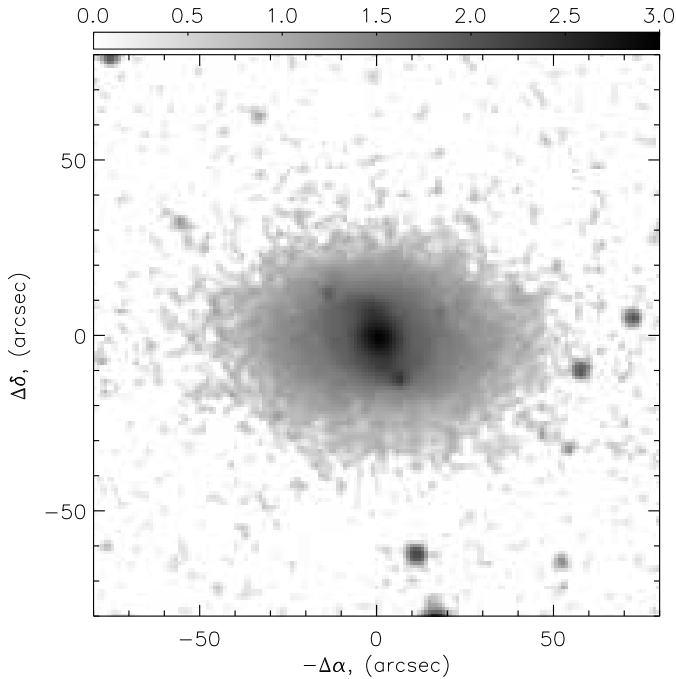}
\caption{ Image of NGC 7177 from 2MASS data, the sum
of three infrared bands J + H + K$_s$. The intensity scale
is logarithmic, with an arbitrary zero point; north is at the
top and east is to the left.}
\end{figure}

The rotation of NGC 7177 was investigated more than once by the method of optical long-slit
spectroscopy and also in the 21 cm HI line with the WSRT radio interferometer
(Rhee and Van Albada 1996).
In the optical wavelength range, the stellar rotation was measured by
Simien and Prugniel (2002) at a position angle of 11$^{\circ}$ and the ionized-gas
rotation was measured by M\'{a}rquez et al. (2002) at a position angle
of 90$^{\circ}$. Both line-of-sight velocity curves rapidly reach a plateau
as one steps from the center. The stellar velocity dispersion measurements by
Simien and Prugniel (2002) show that the stellar disk is dynamically
rather cold ($\sigma_{*,los} <$ 50 km s$^{-1}$) at the distance of $\sim$20$''$
from the center. Thus, neglecting the correction for the asymmetric drift due to
the stellar disk being dynamically cold and assuming
the gas and stars to rotate in the same plane, we can determine the orientation
parameters of the rotation plane by confronting the two rotation curves. Assuming,
following Erwin (2005), that the inclination of the plane to the line of sight
is 48$^{\circ}$, we find the orientation of the dynamical line of nodes:
PA$_{dyn}$ = 80$^{\circ}$. This is in good agreement with the orientation of
the photometric major axis (PA$_0$ = 83$^{\circ}$, Erwin 2005), which confirms our
assumption about a general circular rotation of NGC 7177 outside the visible bar.
The rotation velocity of the galaxy over the inner stellar disk estimated for these
orientation parameters is about 185 km s$^{-1}$.

In this paper, we investigate the structure, stellar population, and star formation history
in the central region of NGC 7177 based on the observational data that we obtained with the Multi-Pupil Fiber Spectrograph of the 6-m telescope. According to present views, it is in
the central regions of disk galaxies that the main consequences of secular evolution, in particular, the results of the matter redistribution in the disk on timescales of 1 –- 5 Gyr governed by the effect of a nonaxisymmetric bar potential, should be sought.

\section*{ OBSERVATIONS AND DATA REDUCTION}
We investigated the central region of NGC 7177 with the Multi-Pupil Fiber Spectrograph (MPFS) of
the 6-m Special Astrophysical Observatory (SAO) telescope in September 2008. The MPFS (Afanasiev
et al. 2001) takes simultaneous spectra of 256 spatial elements arranged in the form of 16 x 16 square lenses array with a scale of 1 arcsec per spaxel. The galaxy was observed twice: on September 5 in the green spectral range (4200 –- 5600 \AA) at a seeing (spatial resolution) of 1$''$.5 and on September 7 in the red spectral range (5800 -– 7200 \AA) at a seeing
better than 1$''$.2. Numerous emission lines of ionized gas ([OI], [OII], [OIII], [NII], [SII], \Ha, and \Hb) and absorption features typical of the old stellar population in galaxies are within the spectral ranges studied. The spectral resolution was about 3 \AA. The detector was a CCD EEV42-40 (2048 x 2048 pixels).

 For the primary observational data reduction, we used a software package developed at the SAO and running
in the IDL environment (Afanasiev et al. 2001). This package used the comparison spectrum of a gas-filled
lamp to calibrate the wavelength scale and the exposure of a flat-field lamp and the twilight sky spectrum
to correct the data for vignetting and different microlens transmissions.

The array (cube) of galactic spectra in the green spectral range was designed to measure
the H$\beta$, Mgb, Fe 5270, and Fe 5335 absorption lines over the field using the recipes of the well-known Lick system (Worthey et al. 1994), the so-called Lick indices. The indices expressed as the equivalent widths of the H$\beta$, Mgb, Fe 5270, and Fe 5335 absorption lines were calculated
from each spectrum and were then collected into ``maps''. Since for these strong absorption lines there exist detailed calculations based on evolutionary synthesis models for an old stellar population (see, e.g., Worthey 1994; Thomas et al. 2003), the mean parameters of the stellar population can be estimated by comparing the observed Lick indices with the
model ones. We also constructed two-dimensional line-of-sight velocity and velocity dispersion fields for
stars in the central galactic region. For this purpose, after the continuum subtraction and the transformation
to a logarithmic wavelength scale, the spectrum of each spatial element was cross-correlated with the
spectra of the defocused (to fill the entire MPFS field of view) G9 subgiant star HD 188512 observed on
the same night and with the same instrumentation as the galaxy. The observations of NGC 7177 both in the
red spectral range ((5800 -– 7200 \AA), based on \Ha and [NII] $\lambda$6583), and in the green one (4200 -– 5600 \AA),
containing [OIII] $\lambda$5007, were used to analyze the ionized gas kinematics. We used a Gaussian fitting of
the profiles for the main emission lines to construct the distributions of their surface brightnesses, line-of-
sight velocities, and widths.

 According to our estimates, the accuracy of individual line-of-sight velocity measurements is about
10 km s$^{-1}$ and the accuracy of determining the equivalent widths of absorption lines at the galactic center
is $\sim$0.15 \AA. We checked the spectrum linearization accuracy and the zero point of the measured velocities
based on the night-sky [OI] $\lambda$5577 \AA ~and [O I]$\lambda$6300 \AA ~lines.

 NGC 7177 has been repeatedly observed with a high spatial resolution on the Hubble Space Telescope
(HST) imaging cameras; we retrieved these data from the HST archive and used them in our
analysis. On July 7, 2001, the galactic center was imaged with the WFPC2 camera through the \emph{F}606\emph{W}
and \emph{F}814\emph{W} filters as part of a program aimed at searching for supernova progenitors (P.I.S. Smartt,
PropID 9042). On September 19, 1997, the central region of NGC 7177 was observed with the
NICMOS2 infrared camera through the \emph{F}110\emph{W} and \emph{F}160\emph{W} filters as part of M. Stiavelli's program
(PropID 7331). The color maps constructed from these data are similar to the \emph{R -– I} and \emph{J -– H} colors in
the optical and infrared ranges, respectively. We will use these maps in discussing the morphology of the
dust distribution in the central region of NGC 7177.

\section*{IONIZED GAS KINEMATICS AND MORPHOLOGY}

Using our two-dimensional spectroscopy data for the central region of NGC 7177, we constructed maps
of the intensity distribution for the strongest emission lines. They are presented in Fig. 2 in arbitrary
units along with the ``green'' and ``red'' continuum maps. The difference between the surface
brightness distributions of the hydrogen and low excitation (nitrogen and sulfur) lines is immediately
apparent: whereas the nitrogen and sulfur emission is the strongest in the galactic nucleus, the \Ha emission
concentrates mainly in three compact emission-line regions located on different sides of the galactic center
at distances from 1$''$.5 to 2$''$. Interestingly, the high excitation [OIII] $\lambda$5007 \AA ~line ``traces'' the same three compact regions where the Balmer emission peaks are observed. It should be noted that the central region of
NGC 7177 was previously mapped through a narrow filter centered at the \Ha emission line, and the galaxy
map in this emission line was published by D\'{i}az et al. (2000) and S\'{a}nchez-Portal et al. (2000). Our
data morphologically agree with the published ones: the same three compact HII regions that surround
the nucleus of NGC 7177 are seen in Fig. 1 from D\'{i}az et al. (2000) and Fig. 7 from S\'{a}nchez-Portal
et al. (2000). Figure 3 shows the [N II] $\lambda$6583/\Ha and [OIII] $\lambda$5007/\Hb ratios for the various galactic
regions that we separated. As is well known, the \Ha/[NII] $\lambda$6583 ratio can serve as a diagnostic of
the ionized gas excitation mechanism: this ratio is larger than unity when the gas is excited by radiation
from young stars, while active nuclei exciting the gas by a nonthermal source and shocks produce a
stronger nitrogen line (see, e.g., Veilleux and Osterbrock 1987). In Fig. 3, where we compare the ratios
of the emission line fluxes measured in five compact regions with the theoretical calculations by Kewley
et al. (2001), we see that the [NII] $\lambda$6583/\Ha ratio is large enough for the shocks or the active nucleus
to be considered the prevailing gas excitation mechanism only in the nucleus of NGC 7177 itself. In the
remaining circumnuclear knots and on the periphery of the region being investigated, northward and
southeastward of the nucleus, the gas excitation is diagnosed either as ``pure'' ongoing star formation or
as boundary excitation between an HII region and a shock. Obviously, we see ongoing star formation in
the regions arranged in the form of a patchy ring at 2$''$ from the center of NGC 7177 and further southward, while
the gas excitation mechanism in the galactic nucleus itself bears no relation to young stars.

 Figure 4 presents kinematical information for the central part of NGC 7177: the maps of the ionized gas
line-of-sight velocities measured from the [NII] $\lambda$6583 emission line, the stellar
line-of-sight velocities, and the velocity dispersions of gas clouds and stars. The stellar component shows
a cylindrical rotation within the investigated region whose dynamical major axis, PA$_{0,kin}$ = 91$^{\circ}$, nevertheless,
deviates noticeably from the orientation of the line of nodes of the galactic symmetry plane
(PA$_{0,lon}$ = 80$^{\circ}$). Bearing in mind that the galactic bar is oriented at PA$_b$ = 13$^{\circ}$ -– 18$^{\circ}$ (de Jong 1996), i.e., the dynamical major axis of the stellar component deviates from the orientation of the line of nodes in a
direction opposite to the deviation of the isophote orientation, it can be surmised that we see the influence
of a nonaxisymmetric bar potential on the regular stellar rotation. The kinematic behavior of the ionized
gas differs significantly from that of the stars. The isolines of the gas line-of-sight velocity field turn noticeably
within our field of view: the orientation of the kinematic major axis of the ionized gas changes from
PA$_{0,kin}$ = 109$^{\circ}$ at the very center of NGC 7177 to PA$_{0,kin}$ = 80$^{\circ}$ at distances from the center larger than 4$''$. Bearing in mind the orientation of the line of nodes of the global galactic disk (PA$_{0,lon}$ = 80$^{\circ}$), we may conclude that the ionized gas line-of-sight velocity distribution within R = 4$''$ is most likely affected by
radial motions. Thus, one might expect the gas inflow toward the galactic center in the bar potential.

\begin{figure*}
\hspace {2cm}
\includegraphics {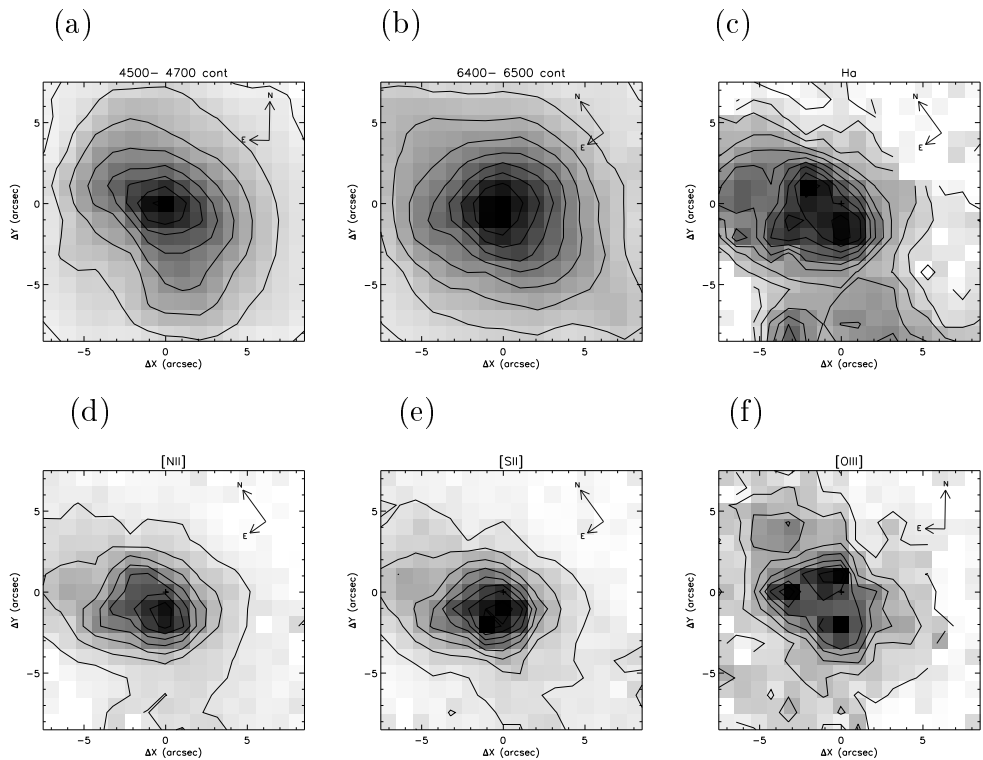}
\caption{Continuum and emission-line surface brightness distributions from the MPFS data: (a) continuum in the wavelength
range 4500 -– 4700 \AA, (b) continuum in the wavelength range 6400 -– 6500 \AA, (c) \Ha, (d) [NII] $\lambda$6583 \AA, (e) [SII] $\lambda$6716 \AA, and (f) [OIII] $\lambda$5007 \AA. The fluxes are normalized in arbitrary units.}
\end{figure*}

%[width=\hsize]
%[width=17cm]
\begin{figure*}
\hspace {2.5cm}
\includegraphics  {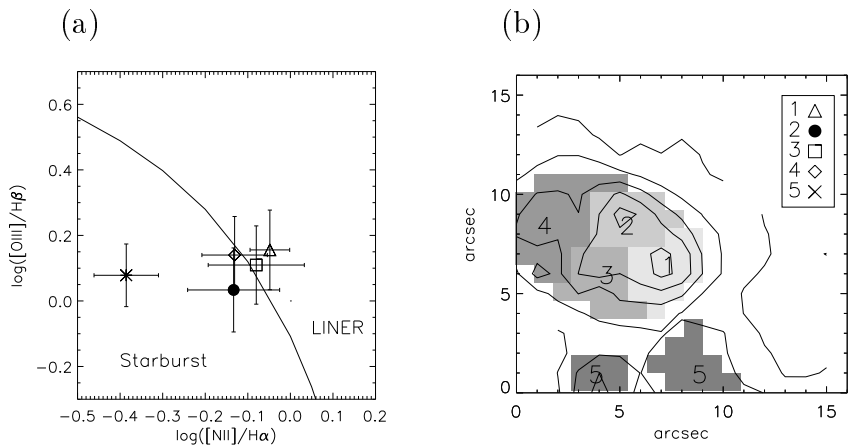}
\caption{(a) Ionization diagnostic diagram for compact regions at the center of NGC 7177; the boundary between the gas
excitation by young stars and active nuclei (shocks) is shown according to the model calculations by Kewley et al. (2001);
panel (b) corresponding to the MPFS field of view shows a mask that separates out the knots being studied with isophotes in
the \Ha emission line superimposed on the map.}
\end{figure*}

\begin{figure*}
\hspace {1.5cm}
\includegraphics [width=15cm] {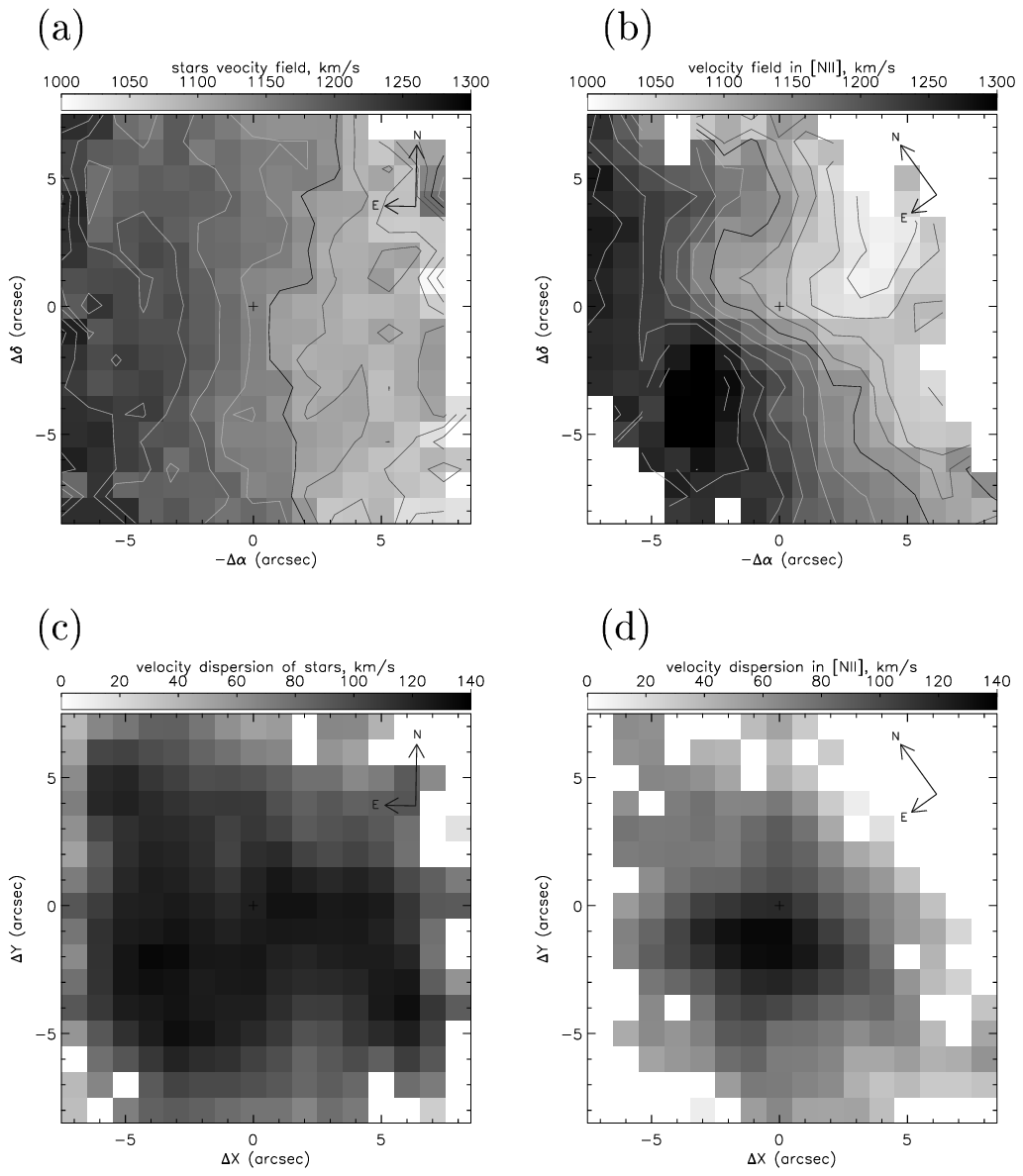}
\caption{Kinematical maps for NGC 7177 from the MPFS data: (a) stellar line-of-sight velocity field (isovelocities + shades of gray), the
line-of-sight velocities are given in km s$^{-1}$; (b) ionized gas line-of-sight velocity field (isovelocities + shades of gray) measured from the [N II] $\lambda$6583\AA ~emission line; (c) the shades of gray represent the stellar velocity dispersion; (d) the shades of gray represent the velocity dispersion of ionized gas clouds estimated from the [N II] $\lambda$6583\AA ~line width.}
\end{figure*}

The stellar velocity dispersion over the field being investigated varies insignificantly between 90 and
140 km s$^{-1}$, showing an overall elongation of the region of maximum in the bar direction; such a behavior
is in agreement with the predictions of dynamical models for the evolution of stellar bars (Vauterin and
Dejonghe 1997). The local stellar velocity dispersion minima can be felt in two of the three central star forming
regions; this is natural, given the ``cooling'' influence of the young stars that have recently been
formed out of a dynamically cold gas on the dynamics of stellar systems. The emission line widths are the highest in the galactic nucleus -— here, the velocity dispersions of the stars and gas clouds are comparable.
Outside the nucleus, the gas is a dynamically cold subsystem.

\begin{figure*}
\hspace {3cm}
\includegraphics[width=10cm]{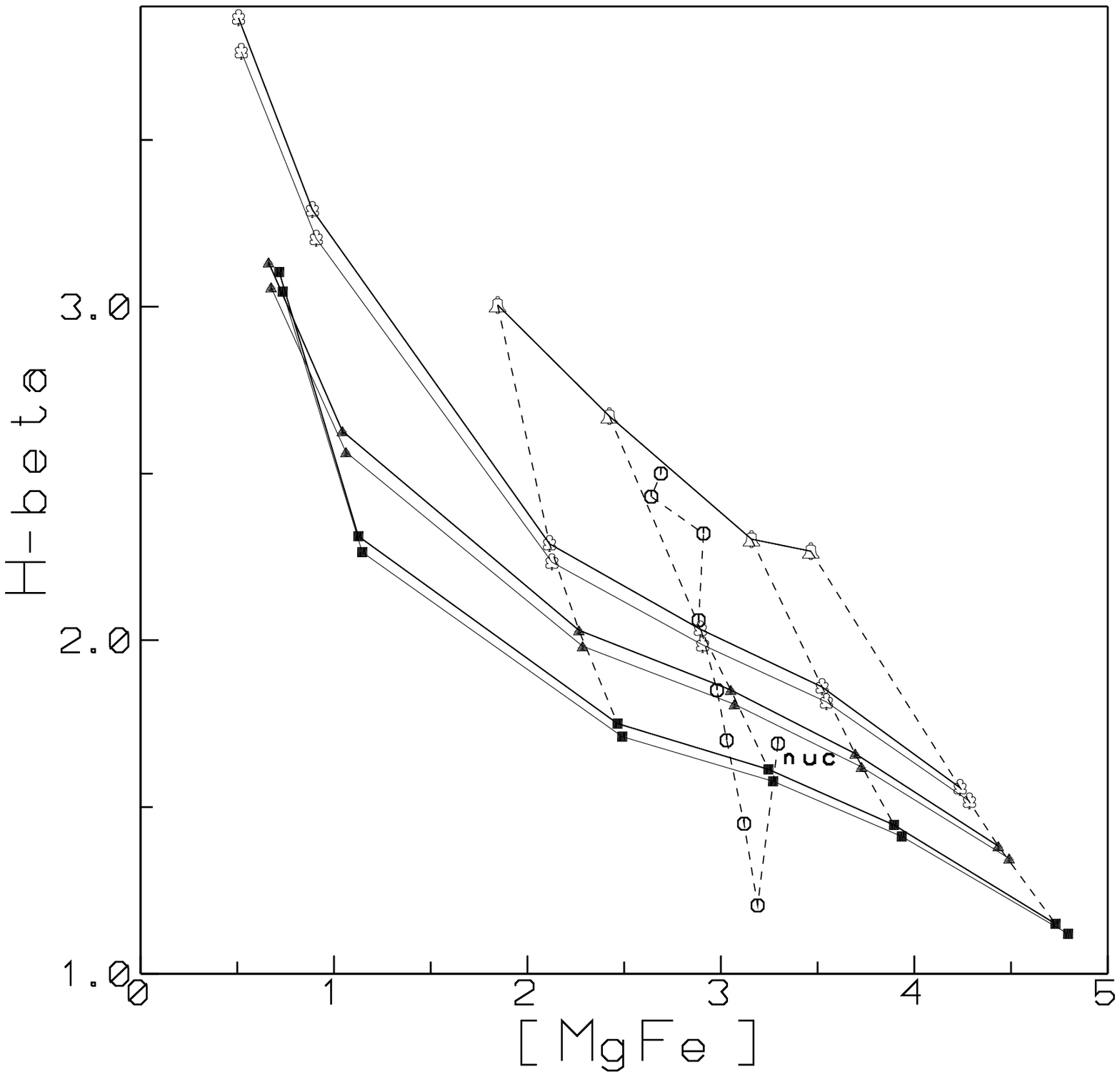}
\caption{ Diagnostic diagram confronting the combined index of metal absorption lines (see the text) and the index of the
Balmer H$\beta$ absorption line to determine the mean age of the stellar population. The large circles joined by the broken line
represent the azimuthally averaged measurements of the indices in NGC 7177 taken along the radius at 1$''$ steps; the galactic nucleus (r = 0$''$) is marked by nuc. The measurements of the
H$\beta$ index were corrected for the emission as described in the text. The observational data are compared with the models from Thomas et al. (2003) for two magnesium-to-iron abundance ratios, [Mg/Fe] = 0.0 (gray polygonal line) and +0.3 (black polygonal line). The model sequences of equal age (solid lines) correspond to T = 2, 5, 8, 12 Gyr (from top to bottom), while the model sequences of equal metallicity (dashed lines) correspond to [Z/H] = +0.67, +0.3, 0.0, -0.33 (from right to left; for ages of 5 -– 12 Gyr, metallicities of -1.35 and -2.35 are also marked by separate symbols).}
\end{figure*}

\section*{THE AGE OF THE STELLAR POPULATION}

We estimated the stellar population parameters at the center of NGC 7177 and their variations along
the radius using the Lick indices H$\beta$, Mgb, and $\langle$Fe$\rangle$ $\equiv$ (Fe 5270 + Fe 5335)/2. To keep a constant
signal-to-noise ratio with increasing distance from the nucleus, we summed the MPFS spectra of the
spatial elements in rings centered on the nucleus with radii of 1$''$, 2$''$, . . . , 8$''$ and measured the indices
in the summed spectra. For such a procedure, the accuracy of the indices is approximately
the same for all rings; at our signal-to-noise ratio, it is approximately equal to 0.15 \AA. Comparison of
the indices Mgb and $\langle$Fe$\rangle$ $\equiv$ (Fe 5270 + Fe 5335)/2 revealed a moderate magnesium overabundance with
respect to iron at the center of NGC 7177: [Mg/Fe] $\approx$ +0.1. To avoid biased age estimates for the stellar
population unavoidable at a nonsolar heavy-element abundance ratio if H$\beta$ is compared with the index
of anyone element, we use a combined metal index [MgFe] $\equiv$ (Mgb$\langle$Fe$\rangle$)$^{1/2}$ in our diagnostics. In
Fig. 5, we compare our measurements of H$\beta$ and [MgFe] $\equiv$ (Mgb$\langle$Fe$\rangle$)$^{1/2}$ with the models by Thomas
et al. (2003) for [Mg/Fe] = 0.0 and +0.3. We see that the age estimates depend weakly on the magnesium-to-iron ratio when the combined metal index is used. Figure 5 plots the values of
H$\beta$ corrected for the emission that is clearly seen over the entire field of view in the central part of NGC 7177.

 We made this correction by using our measurements in the red spectral range for the equivalent
width of the \Ha emission line. In our view, this is the most reliable method, because there exist
physical models for the ratio of the \Ha to \Hb Balmer emission-line intensities and because the
intensity of the \Ha emission line is always several times higher than that of the \Hb one, while the
equivalent width of the \Hb absorption line is generally of the order of or larger than that of the \Ha one.
The \Ha/\Hb emission-line ratio is the smallest, of 2.5 -– 2.7 (Burgess 1958), in the case of hydrogen excitation
(ionization) by UV radiation from young massive stars (HII-region excitation type). In active
nuclei and shock fronts, this ratio is more than three due to the excitation mechanism itself -— by a shock or
photoionization by radiation with a power-law spectrum; a noticeable presence of dust in the emitting
region also increases this ratio. The observational statistics including the spectra of spiral galaxies with
various types of excitation gives a sample-averaged ratio EW(\Hb emis) = 0.25EW(\Ha emis) (Stasinska
and Sodre 2001). As we ascertained in the previous section, individual star-forming regions are observed
at the center of NGC 7177 against the general background of shock excitation of the emission-line
spectrum. In the red-range spectra co-added in the rings, the nitrogen emission line is stronger
than the \Ha emission line at all distances from the center. Therefore, correcting the
index H$\beta$ for the emission in the central region of NGC 7177, we will take
EW(\Hb emis) = 0.25EW(\Ha emis) in all rings within 8$''$ of the center.

We see on the diagnostic diagram in Fig. 5 that the galactic nucleus is old, with a mean stellar population
age of $\sim$ 10 Gyr. As one recedes from the center, the mean stellar population age decreases and reaches a
constant value of $\sim$ 2 -– 3 Gyr starting from a radius of $\sim$ 6$''$. The mean stellar metallicity is almost constant
along the radius, being approximately solar.

\section*{DISCUSSION}

The morphology of the central region in NGC 7177 has been repeatedly studied using data from the
HST, which provided the galaxy images in various spectral bands, from the ultraviolet to near-infrared
ones, with a high spatial resolution. The central region of the galaxy is rich in dust, whose distribution
at the center of NGC 7177 is characterized in the literature as ``extremely chaotic''(Gonz\'{a}lez Delgado
et al. 2008). Figure 6 presents the \emph{F}110\emph{W} -– \emph{F}160\emph{W} infrared color map for the central region of NGC 7177 that we constructed from NICMOS/HST data (the spatial resolution is 0$''$.2) -- this color band is close
to $(J-H)$. The map was turned so as to correspond in orientation to the field exposed for NGC 7177 with
MPFS/BTA in the red spectral range. We see that the dust at the center of NGC 7177 is optically thick:
it is clearly traceable in the form of red lanes southeastward of the nucleus even in the infrared color.
The \emph{F}606\emph{W} -- \emph{F}814\emph{W} (WFPC2/HST) color map (not shown here) demonstrates the same morphology.
However, we would not classify the dust distribution at the center of NGC7177 as chaotic; on the contrary,
the dust lanes extend quite orderly along the southern ``leading'' edge of the bar. As the center is approached,
within a radius of $\sim$ 4$''$, the dust lane turns almost though a right angle and crosses the nucleus at
PA $\approx$ 70$^{\circ}$. Such a morphology of the dust distribution is well reproduced in the dynamical models of the
``response'' of a gaseous disk to a rotating bar if the bar has a double inner Lindblad resonance and
is explained by the structure of the shock fronts in the gas flowing on the bar (Athanassoula 1992).
However, in this case, only ``half'' of the bar shock fronts is seen in NGC 7177; a symmetric extension of
the dust pattern northward of the nucleus should be expected in the theory. If we have taken
into account that the galactic
disk is inclined approximately by 50$^{\circ}$ to the plane of the sky and that the southern side of the disk is
nearest to us (in order that the southeastern edge of the bar be the leading one), then the absence of
the northern dust lane on the color maps can mean only one thing: the bar in NGC 7177 is vertically
thick and screens the shock front in the thin gaseous disk northwestward of the nucleus. A photometric
analysis of the surface brightness profile shows that the bulge in the galaxy is compact and it dominates
over the disk at galactocentric distances R $<$ 8$''$ (Baggett et al. 1998) –- R $<$ 15$''$ (Graham 2001); i.e.,
the bar in NGC 7177 with a semimajor axis of only 10$''$ -– 13$''$ (de Jong 1996; Erwin 2005) is a structure
of the bulge rather than the disk and, in some sense, it is the bulge itself or, more precisely, a ``pseudobulge''
whose formation mechanism can be dynamical bar heating perpendicular to the disk plane (Chung and
Bureau 2004).

 According to the model results by Athanassoula (1992), the change in the orientation of the
dust lanes as the center is approached is evidence for the presence of a double inner Lindblad resonance
in the bar. As we made sure above, the \Ha emission is located around the nucleus in the form of knots
at equal distances of 1$''$.5 -– 2$''$, which can be characterized as an inhomogeneous nuclear star formation ring.

\begin{figure*}
\hspace {3cm}
\includegraphics[width=12cm]{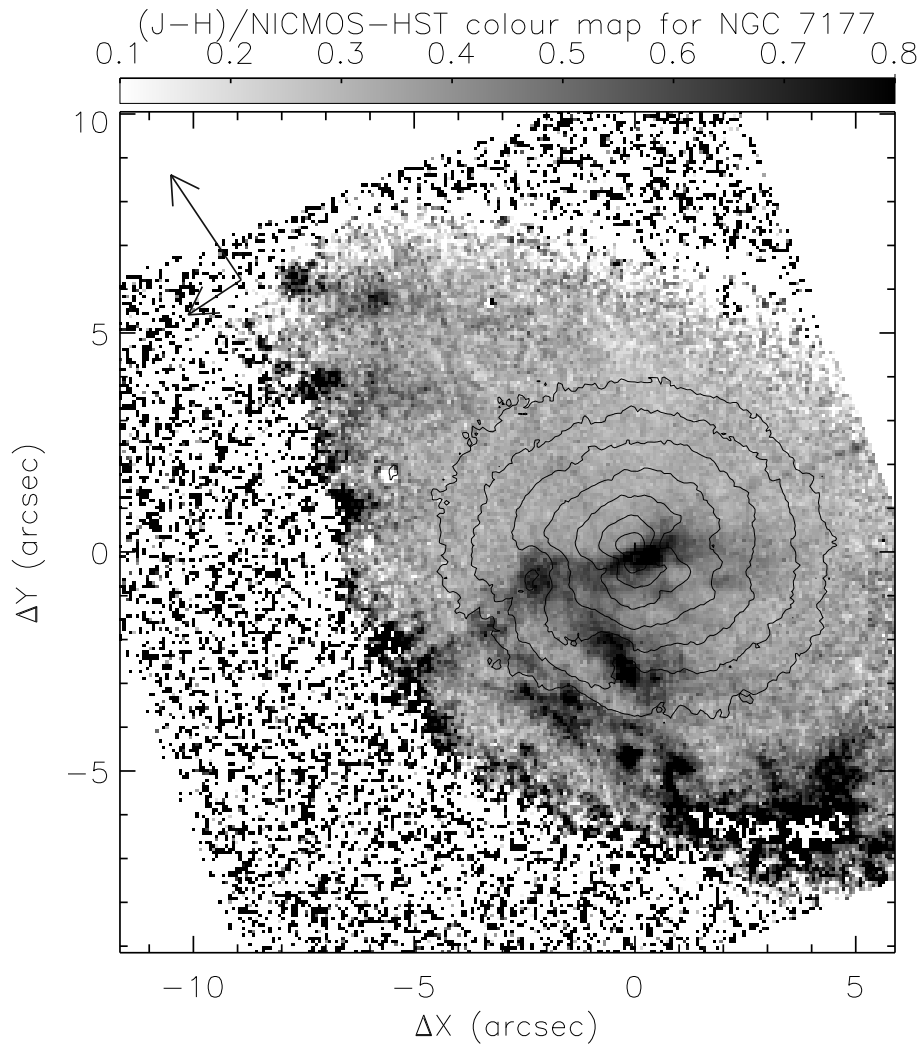}
\caption{\emph{F}110\emph{W} -– \emph{F}160\emph{W} \emph{(J –- H)} color map for the central region of NGC 7177 from the NICMOS/HST data. The color zero point is arbitrary.}
\end{figure*}

The nuclear starforming rings are also usually associated with the inner Lindblad resonances of the
bars (Buta and Crocker 1993) and theoreticians place them between the inner inner and outer inner Lindblad
resonances (Heller and Shlosman 1996). Can we estimate the possible locations of the inner Lindblad
resonances in NGC 7177? As we mentioned in the Introduction, the galactic rotation curve was measured
by M\'{a}rquez et al. (2002) from the ionized gas; in the Introduction, we also refined the orientation of
the line of nodes for NGC 7177. Having calculated the circular rotation curve in the galactic plane from
the data by M\'{a}rquez et al. (2002), we made sure that it demonstrates a local maximum at a galactocentric
distance R $\approx$ 5$''$. Figure 7 presents this rotation curve in angular units of km s$^{-1}$ arcsec$^{-1}$ (solid curve)
and the dependence on radius for $\Omega - \kappa$/2 (dashed curve). The intersection of the latter curve with the
constant at the level of the bar pattern speed gives the position of the inner Lindblad resonance. If there
is a fast bar in NGC 7177 typical of early-type spiral galaxies (Rautiainen et al. 2008), then the ratio of the
corotation radius to the bar radius is 1.0 -– 1.4. The upper boundary of this ratio places the inner inner
Lindblad resonance in NGC 7177 at R $\approx$ 2$''$ precisely where the star formation ring is observed. So, the
set of kinematic and morphological data is consistent with the standard model picture of the gas motion and
the dust distribution between the two inner Lindblad resonances at the center of NGC 7177.

 The mean stellar age inside the star formation ring in the galactic nucleus is $\sim$ 10 Gyr. Outside the
star formation ring, at a distance R = 6$''$ -- 8$''$ from the nucleus, the mean age of the stellar population
is $\sim$ 2 Gyr. The dynamical and morphological signatures suggest that the bar in NGC 7177 is old:
it has managed to thicken in z coordinate and the stellar velocity dispersion in the galactic center is at
a maximum, which is also indicative of an advanced stage of dynamical evolution. If the bar lifetime and
the mean stellar population age at the outer inner Lindblad resonance are comparable, then it can be
surmised that the star formation ring has migrated radially inward in the last 1 -– 2 Gyr; such a behavior of
the nuclear star formation rings is actually observed in barred galaxies (see the Introduction in van de Ven
and Chang (2009)) and is predicted by various dynamical models of nuclear resonant rings (Regan and
Teuben 2003; van de Ven and Chang 2009). This wave of star formation (and gas concentration) has
not yet reached the nucleus and as yet there have been no chances for the supermassive black hole at the center of NGC 7177, if it exists, to become a truly active nucleus.

\begin{figure*}
\hspace {3cm}
\includegraphics[width=10cm]{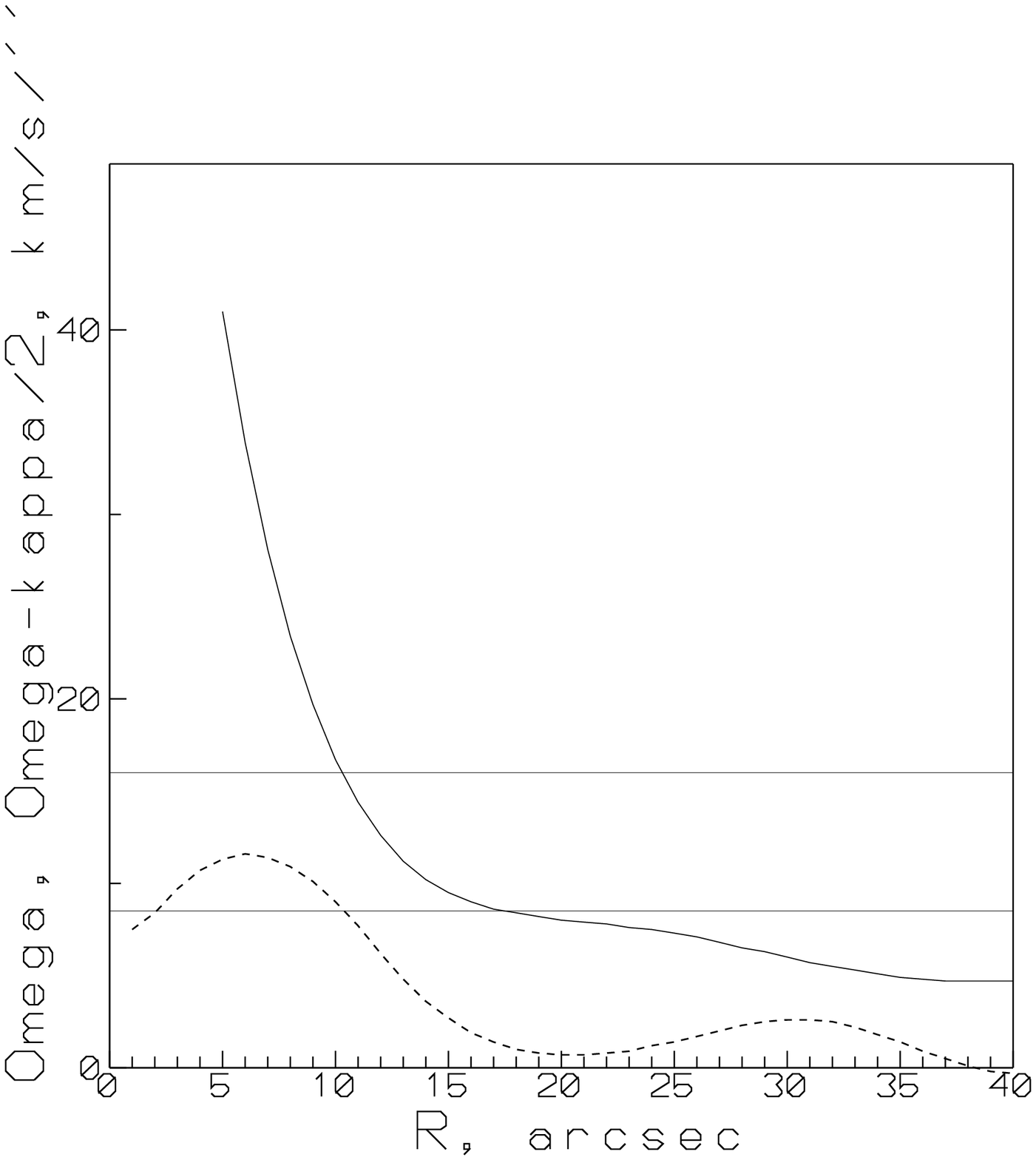}
\caption{Rotation curve in the central region of NGC 7177 based on the data from M\'{a}rquez et al. (2002) and positions of the inner Lindblad resonances; the interval for the angular velocity of the bar denoted by the thin horizontal solid lines is taken according to the assumption that the bar in NGC 7177 is fast.
}
\end{figure*}

\begin{acknowledgements}
We wish to thank A.V. Moiseev, a leading researcher from the Special Astrophysical Observatory
of the Russian Academy of Sciences, for support of the MPFS observations on the 6-m telescope.
We used data from the NASA/ESA Hubble Space Telescope operated by the Association of Universities
for Research in Astronomy under contract with NASA, NAS 5-26555. During our work, we relied
on the means of HYPERLEDA, the Lyon–Meudon Extragalactic Database, provided by the LEDA team
at the Lyon CRAL Observatory (France) and the NASA/IPAC database (NED) operated by the Jet
Propulsion Laboratory of the California Institute of Technology under contract NASA. The study of the
star formation history in the central regions of galaxies with various types of global stellar disks was
supported by the Russian Foundation for Basic Research (project no. 07-02-00229a) and the observational
study of the gas kinematics at the centers of galaxies was supported by grant no. 09-02-00870a.

\end{acknowledgements}

\vspace {2cm}

ASTRONOMY LETTERS Vol. 36 No. 5 2010
\end{document}